\documentclass[aps,prc,showpacs,preprint]{revtex4-1}
\usepackage{graphicx}
\usepackage[section]{placeins}
\begin{document}
\title{Hamilton-Jacobi Approach to the Quantization of Classically Non-Separable but Integrable Two-Dimensional Systems: the Role of the Classical Caustics}
\author{M. Fusco Girard}
\affiliation{Department of Physics ``E.R. Caianiello",\\University of Salerno \\and \\Gruppo Collegato INFN di Salerno,\\Via Giovanni Paolo II, 84084 Fisciano (SA), Italy\\
electronic address: mario@sa.infn.it}

\begin{abstract}
The quantization method based on the quantum Hamiltonian Jacobi equation, is extended to two-dimensional non-separable but integrable Hamiltonians. It is shown that each wave function for those systems corresponds to a well-defined family of classical trajectories, enveloped by a caustic. The energy eigenvalues and the values of the wave functions on the caustic are obtained by solving the 1-dim quantum Hamilton Jacobi equation of the caustic' arcs. Results are in good agreement with those obtained by usual methods.
\end{abstract}

\maketitle

A non-separable two-dimensional Hamiltonian system can be quantized in various analytical or numerical ways: e.g., perturbative methods, diagonalization of the truncated Hamiltonian matrix, variational methods, or else direct numerical integration of the Schr\"odinger equation [1]. However, no one of these methods allow to study the semi classical limit, i.e. what happens to the relevant quantum quantities when $h \to 0$: in fact, all they overlook the profound links between the quantum system and the corresponding classical trajectories.

Recently a method, which enables to investigate these problems for one dimensional or separable cases, has been presented. This method is the exact version of the WKB semi classical approximation, and is based on the determination, numerical  [2, 3] or analytical [4], of solutions of the Quantum Hamilton-Jacobi Equation (QHJE), which is the equation for the complex phase of solutions of the Schr\"odinger Equation (SE) [5, 6]. This approach is completely in the framework of the usual Copenhagen interpretation of the quantum mechanics. It gives the same results of the SE for the energy eigenvalues and the wave functions, but moreover, it allows to fully investigating the semi classical limit $h \to 0$, where instead the SE loses its significance. The aim of the present paper is to extend the method to the case of two-dimensional non-separable but integrable, time independent Hamiltonian systems. Being the total energy E conserved for these systems, it suffices to consider only time independent quantities.

In every spatial dimensions, the QHJE supplies the link between the quantum and classical mechanics. Indeed, by denoting with $\vec q$ the set of spatial coordinates, the complex phase of every quantum wave function $\psi( \vec q, E)$ is a solution $W_Q (\vec q, E)$ of the QHJE. This function for $h \to 0$ becomes a solution $W_C(\vec q, E)$ of the Classical Hamilton Jacobi Equation (CHJE), which defines a family of trajectories in the configuration space. In this way, to every wave function $\psi (\vec q, E)$ it corresponds a particular bundle of classical trajectories. $W_C (\vec q, E)$ is the Hamilton's characteristic function [7], also called the abbreviated action [8].  For simplicity, in the following we will omit the word ``abbreviated", so that $W_C(\vec q,E)$  will be named the classical action, and $W_Q(\vec q,E)$  the quantum action. (Usually, this terminology is reserved to the function $S_C(\vec q, t)$, which is the Legendre transform of $W_C(\vec q,E)$).

The 1-dim quantization method presented in [2-4] requires finding the solutions of the QHJE in the various regions of the coordinate axis $x$, separated by the classical turning points, where the potential $U(x)$ equals the total energy E. These solutions $W_Q(x, E)$ are complex inside the classically allowed regions, and purely imaginary outside, and are the phases of solutions of the SE. For a generic value of the energy, these latter are not regular functions of the coordinate, i.e. are not normalizable and everywhere continuous with their first derivative. In order to get a regular quantum wave function, the quantization condition has to be satisfied. This is the same as for the approach based on the SE: a value of the energy is allowed if with this value of the parameter in the QHJE, it is possible to construct a regular wave function, by matching together the solutions of the SE, obtained in the various regions by means of the solutions of the QHJE. The method is exact and gives the energy eigenvalues and the wave functions with great precision. In the limit $h\to 0$, it reduces to the semi classical WKB approximation. As in the latter, in this approach a central role is played by the turning points. In these, the second derivative of the wave function vanishes; they separate the regions where the wave function has an oscillating behavior (the classically allowed regions) from those where it exponentially decreases (the classically forbidden ones).

Let us consider now a two-dimensional non-separable but integrable Hamiltonian. For such a system, the classical trajectories in the phase space belong to invariant tori, whose projection on the configuration space $\vec q = (x, y)$ is circumscribed by caustics' arcs, i.e. envelopes of trajectories' families.
One typical example is given in Fig. [1].

\begin{figure}[!h]
\includegraphics[scale=1]{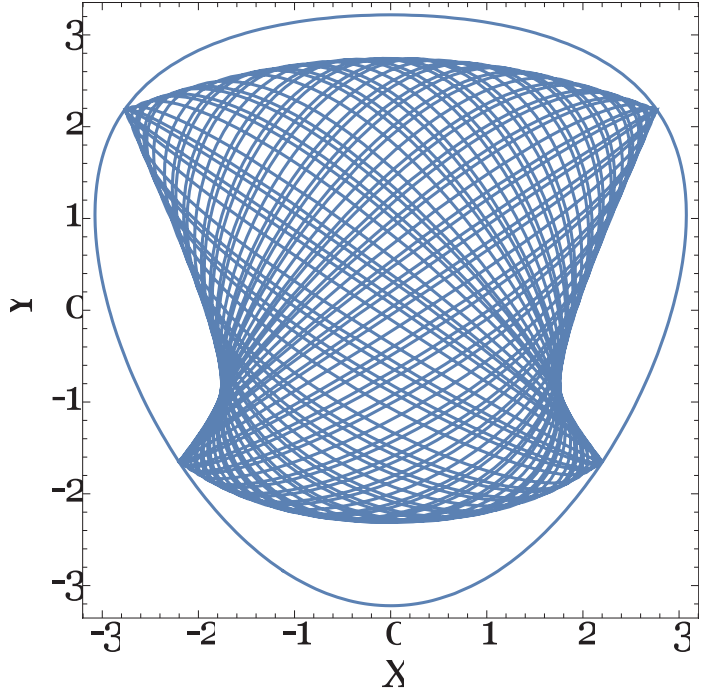}%
\caption{A family of trajectories for the Barbanis system with the Hamiltonian given by Eq. [1]. The frequencies are $\omega_X =1.1$ and $\omega_Y =1.0$, and the coupling constant has the value $\lambda = -0.11$. The energy is $E= 5.18266$.}
\end{figure} 

It refers to the Barbanis system, whose Hamiltonian (with unitary mass) is:
\begin{equation}
H(x,y)={1\over 2} \left ( p_x^2+p_y^2\right) + {1\over 2} \omega_x x^2+{1\over 2} \omega_y y^2 + \lambda x^2 y
\end{equation}   
The caustics' arcs touch in four points the equipotential line $U(x, y) = E$, which delimitates the region of the configuration space allowed to all the trajectories with the same energy E. We will call these points the (caustics') vertices and will label them by an integer $k$, where $k = 1$ for the lower left vertex, and the others follows in clockwise direction. The caustics' arcs are analogously enumerated, starting from the left arc, between vertices 1 and 2.  In the following, we will denote with F the family of the trajectories enveloped by the same caustics; the ensemble of the caustics' arcs will be called the (family's) caustic $C_F$, and we will denote as $\Omega_F$ the region inside $C_F$.

In the generic case, from each trajectories' family it is possible to extract two (as in the case of the figure [1]) or more (when part of the caustic is internal to $\Omega_F$) sub-families SF of non-intersecting trajectories, covering the entire region $\Omega_F$. By choosing one of these sub-families, in each point the momentum $\vec p$ is uniquely defined, a part for its sign. This ambiguity is removed by orienting the trajectories.

Through the relation:
\begin{equation}
 \vec p(x,y,E)={\rm grad}\ W_C(x,y,E)\ ,
 \end{equation}	     
 to this vectorial field it corresponds a particular classical action $W_C(x, y, E)$, which in turn is the limit for $h\to 0$, of a specific quantum action $W_Q(x, y, E)$.

Just like for the one-dimensional case, this latter function is the phase of a solution of the SE which, if E is an energy eigenvalue, is a regular quantum wave function $\psi_{W_Q}(x, y, E)$  (in the following, we will often omit the explicit dependence on the energy E). The construction of $W_C(x, y)$ and the corresponding semi classical wave function was demonstrated [9, 10] for non-separable integrable systems by means of a method based on the solution of a suitable Dirichlet's problem for the time-independent 2-dim CHJE:
\begin{equation}
\left ({\partial W_C(x,y)\over \partial x}\right )^2 + \left ({\partial W_C(x,y)\over \partial y}\right )^2 =2m\left (E-U(x,y)\right)
\end{equation} 

In that case, too, the caustic plays a crucial role, being the boundary for the Dirichlet's problem. Chosen a sub-family SF of non-intersecting trajectories enveloped by the caustic $C_F$ and spanning the region $\Omega_F$, and indicating with $(x_C, y_C)$ the generic point on the caustic, the value of the classical action $w_C(x_C, y_C)$ there is obtained in the following way. Firstly, the caustic is oriented, afterward the form $\vec p  d \vec q$ is integrated along the caustic from a chosen vertex $(x_V, y_V)$ to  $(x_C, y_C)$, where $\vec p$ in each point is the momentum of the trajectory touching the caustic in that point [11]. Finally, the classical action $W_C(x, y)$ in each point $(x, y)$ of $\Omega_F$ is obtained by integrating the same form along the trajectory from the caustic to $(x, y)$ and using  $w_C(x_C, y_C)$ as the initial value:
\begin{equation}
W_C (x,y)=w_C (x_C,y_C)+\int_{(x_C,y_C)}^{(x,y)}\vec p d \vec q
\end{equation}      

The consideration of non-intersecting trajectories avoids the problem of many-valued solutions. $W_C(x, y)$ is a real function inside the region enveloped by the caustic, and can be extended to complex values outside this region, by considering imaginary trajectories, i.e. solutions of the Hamilton's equation in the regions where $U(x, y) > E$. Outside $\Omega_F$, at least one of the components of the momentum $\vec p$ is imaginary, so that the integration of $\vec p d\vec q$ along these trajectories produces a complex or an imaginary action, depending on the case.

As an introduction to the quantization problem, let us consider the Hamiltonian (1) with no coupling, i.e. for $\lambda = 0$, and incommensurate frequencies. This system is equivalent to a pair of non-interacting harmonic oscillators, and $\Omega_F$ is a rectangle. The caustic's arcs are the sides of the rectangle, and the trajectories' family F spans the region inside it. In each point, two trajectories meet, so from F two non-intersecting sub-families can be extracted. By choosing one (SF) of these sub-families and orienting the trajectories, in each point of $\Omega_F$ the momentum $\vec p$ results uniquely defined. On the caustic the normal component of $\vec p$ vanishes, while it is real and different from zero at one side of the caustic, in the region $\Omega_F$, and imaginary at the other side. As for the kinetic energy, near the caustic it has a part tied to the motion along the caustic, while the other one is linked to the motion normal to it. This latter is real inside the caustic, vanishes on it, and becomes negative at the other side. The QHJE is separable and the total quantum action $W_{Q,SF}(x, y)$ is simply the sum of the actions for the two degrees of freedom. The analytical expressions for the two one-dimensional quantum actions are given in [4], and will not be reported here. From these it follows that $W_{Q,SF}(x, y)$  is a complex function in $\Omega_F$.

The wave function is linked to $W_{Q,SF} (x, y)$  by the relation:
\begin{equation}
\psi(x, y) = A \exp \left({ W_{Q,SF} (x, y)\over \hbar }\right )
\end{equation}			
where A is a constant,  and inside $\Omega_F$ it is an oscillating function, while outside is exponentially decreasing (monotonically or oscillating). In Fig. [2], the surface corresponding to the real part $X_{Q,SF} (x, y)$ of  $W_{Q,SF} (x, y)$ in the region $\Omega_F$ for the non-interacting harmonic oscillators is presented (left box), while in the right box the corresponding classical quantity $W_{C,SF} (x, y)$ is plotted.

\begin{figure}[!h]
\includegraphics[scale=0.8]{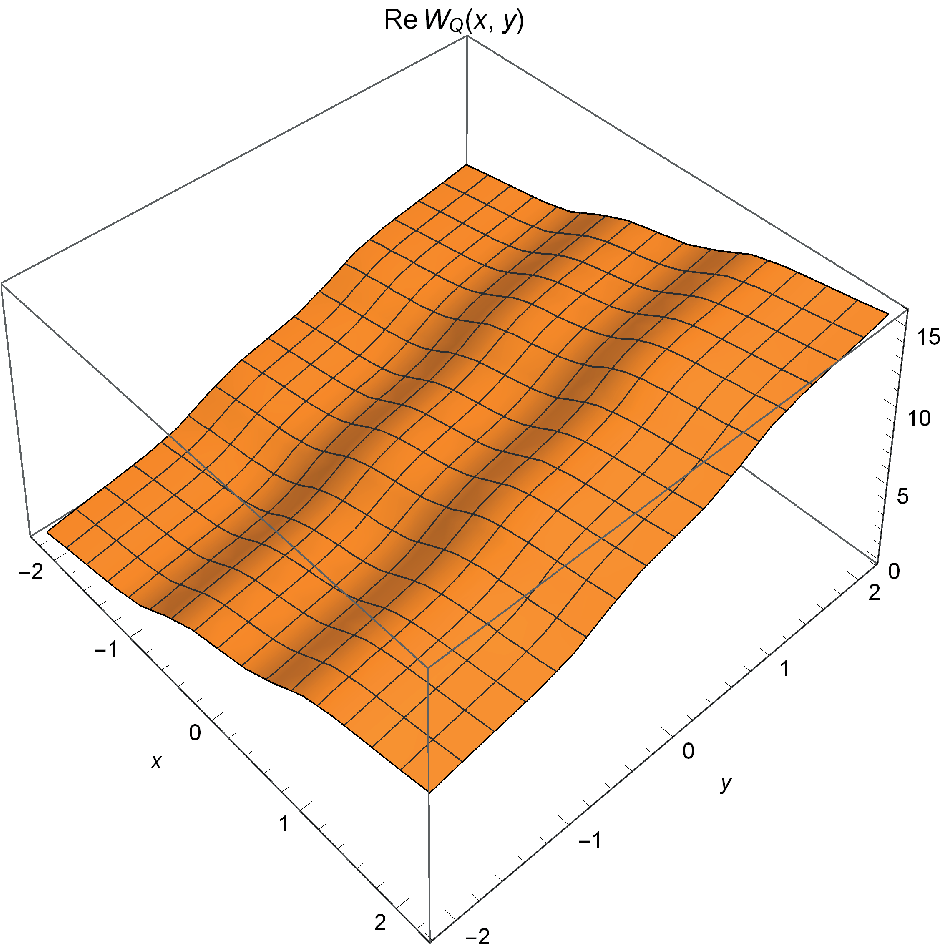}%
\includegraphics[scale=0.8]{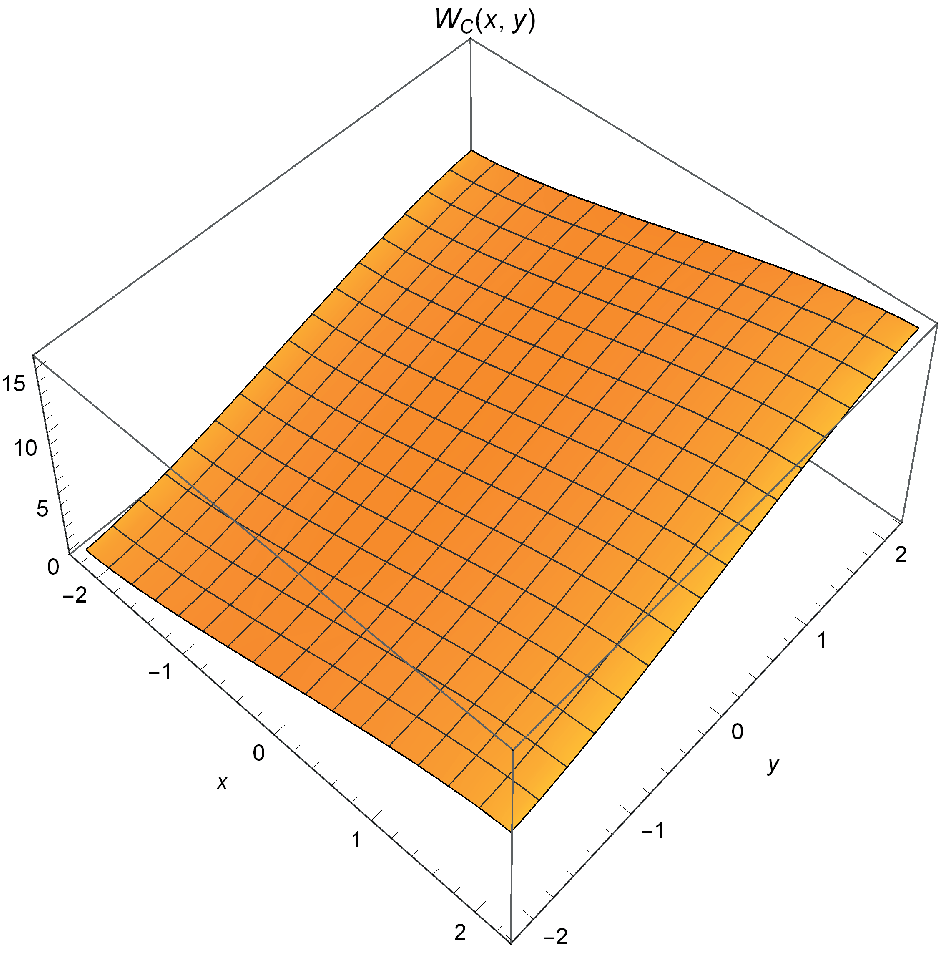}%
\caption{Left: The graph of the real part $X_{Q,SF} (x, y)$ of  $W_{Q,SF} (x, y)$ in the region $\Omega_F$ for a pair of non-interacting harmonic oscillators. Right: for the same system, the graph of the classical action $W_{C,SF} (x, y)$, which is the limit of $X_{Q,SF} (x, y)$. The frequencies and the energy are the same as in Fig. 1.}
\end{figure} 

The two figures refer to the state (2, 2), with the energy $E= 5.18266$. As seen from the figures, the quantum action smoothly oscillates around the classical one. This behavior is the same as for the 1-dim case, and can also be seen in the Fig. [3], where the contour lines for $X_{Q,SF}(x, y)$ and $W_{C, SF} (x, y)$, respectively, are presented.

\begin{figure}[!h]
\includegraphics[scale=0.8]{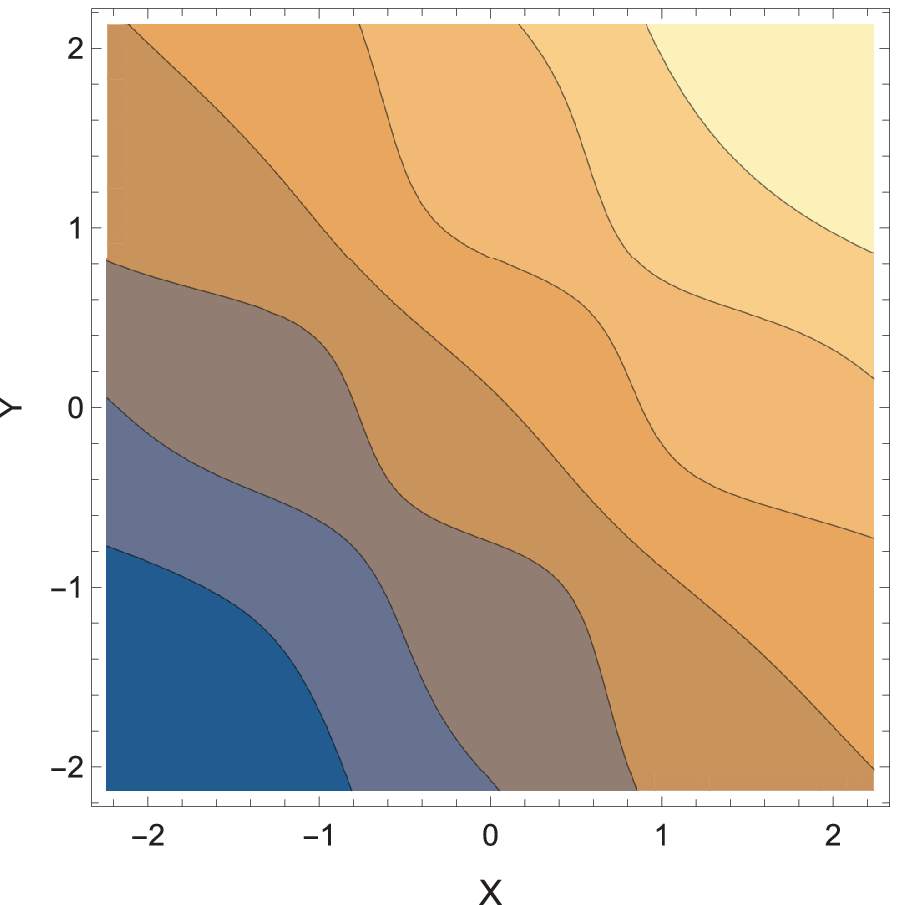}%
\includegraphics[scale=0.8]{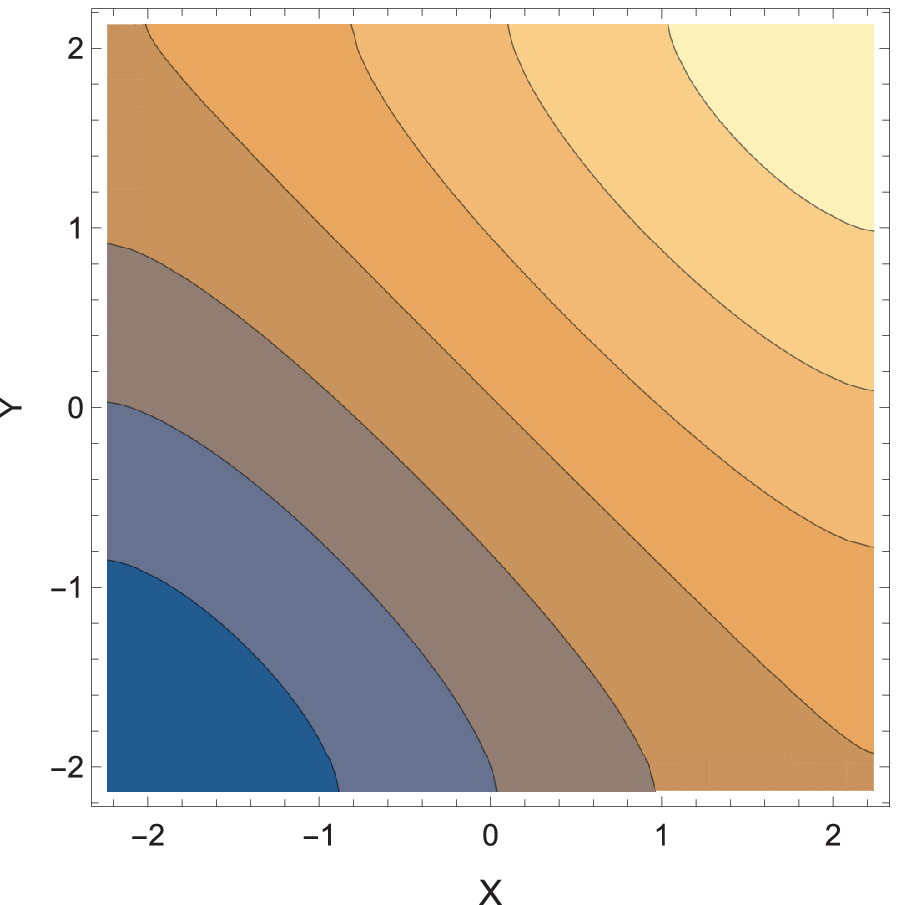}%
\caption{Left: Contour lines in the region $\Omega_F$ for the real part $X_{Q,SF}(x, y)$ of the quantum action $W_{Q,SF} (x, y)$ for a pair of non-interacting harmonic oscillators. Right: Contour lines in the region $\Omega_F$ for the classical action $W_{C,SF} (x, y)$ which is the limit of $X_{Q,SF} (x, y)$ for the same system. Parameters' values as in the other Figs.}
\end{figure} 

The values of the wave function computed by means of Eq. (5) are exactly the same as computed by direct multiplication of the two Hermite polynomial-based solutions for the two separated degrees of freedom. In a more general situation, in which one or both the separated oscillators are not analytically integrable, the quantum actions for the two degrees of freedom can be obtained by numerically integrating the one-dimensional appropriate QHJE on each segment of the caustic between two adjacent vertices, as in [2, 3]. In this way, the quantum action and, if the quantization conditions are fulfilled, the energy eigenvalues and the wave functions for the two separated degrees of freedom can be computed.  This is also true for a generic interacting non-separable but integrable two-dimensional system. Indeed, near the caustic one can introduce a coordinate $\xi$ along the caustic and another $\eta$ normal to it and expand the potential energy at first order in the $\eta$ coordinate. The CHJE, the QHJE and the SE become locally separable, and the same conclusions as before for the momentum, the kinetic energy and the quantum action are then reached. The part depending on $\eta$ of the wave function near the caustic is proportional to an Airy function, and therefore its second derivative normal to the caustic, vanishes on the caustic itself. The same happens at the turning points in the 1-dim case. In two dimensions, hence, the classical caustic plays a role analogous to the turning points in the one-dimensional case, and in particular, it separates the region where the wave function has an oscillating behavior, from those in which it exponentially decreases (monotonically or oscillating).

Therefore, in order to extend to the interacting quantum case the Dirichlet's problem above outlined to construct the classical action, one has to consider a family SF of oriented classical trajectories enveloped by a caustic $C_F$ and spanning the region $\Omega_F$.  In order to put and solve the Dirichlet's problem for the quantum action in $\Omega_F$ one has to preliminarily find the boundary values of $W_{Q, SF}$ on the caustic $C_F$. This requires solving four one-dimensional problems for the QHJE, along each caustic's arc. To this purpose, the numerical procedure exposed in [2, 3] can be applied, with the difference that the equations to solve are to be written for curvilinear arcs. Let us remember that in the rectilinear case, the QHJE can be derived from the SE by looking for solutions in exponential form. Therefore, in order to get the QHJE on a curve, the first step is to write the SE on it. On a caustic's arc labelled with $k$, represented for instance by the function $y=f_k(x)$, the time-independent SE is:
\begin{equation}
-\left[\psi '' (x)-{g'_k(x)  \over g_k(x)} \psi'(x)\right]={2m\over \hbar^2}\left(E-U_k (x)\right) g_k^2(x) \psi(x)\quad ,
\end{equation}	
where the scale factor $g_k(x)$ is
\begin{equation}
g_k(x)=\sqrt{1+f_k'^2 (x)}\quad ,
\end{equation}			
and $U_k(x)$, the 1-dim effective potential energy on the $k$-th arc, is the restriction of the two-dimensional potential energy $U(x, y)$ to the arc $k$:
\begin{equation}
U_k(x) = U(x, f_k(x))
\end{equation}		

The usual rectilinear SE is recovered from Eq. (6), because in that case $f_k'(x) = 0$ so that $g_k(x) = 1$. If the caustic's arc is represented by the function $x=f_k(y)$, it is only necessary to exchange the two variables in the Eq. (6) and following.

Then we search for solutions   $\psi_k (x)$ of Eq.(10) in the form:
\begin{equation}
\psi_k (x)=exp\left ({i\over \hbar} W_{Q,k} (x)\right)\quad .
\end{equation}		
In a classically allowed region, as the $k-th$ caustic's arc itself between two adjacent vertices, the quantum action $W_{Q,k}(x)$ is looked for as a complex function:
\begin{equation}
W_{Q,k}(x) = X_k(x) +i Y_k(x)\quad .
\end{equation}		
By substituting this into the one-dimensional SE (6), and by separating the real and the imaginary parts, one gets the equation (apices denotes derivatives with respect to the argument):
\begin{equation}
-{X'^2_k (x)} +{Y'^2_k (x)} +{Y'_k (x) g'_k(x)\over g_k(x)} -{Y''_k (x)} = \left (-{2m\over \hbar^2} \right)\left(E-U_k(x)\right) g_k^2(x) \quad ,
\end{equation} 	
for the real part, while for the imaginary part:
\begin{equation}
-{2X_k' (x)Y_k' (x)} -{X_k' (x)g_k' (x)\over g_k(x)} +{X_k'' (x)} = 0\quad .
\end{equation}	
The solution of this latter is
\begin{equation}
Y_k(x)=\hbar \log\sqrt{X_k' (x)\over g_k(x)}
\end{equation}		

The last equation in particular shows that when $\hbar \to 0$, the imaginary part of the quantum action on the arc vanishes.
By substituting the last equation into Eq. (11), we get the equation for the real part $X_k (x)$ of the quantum action on the arc $k$:
\begin{eqnarray}
+3 \hbar ^2 g'^2_k(x) X'^2_k (x) + && 4g^2_k(x) X'^4_k (x) - 3 \hbar g^2_k(x) X''^2_k (x) + 2\hbar ^2 g^2_k(x) X_k'(x)  X_k'''(x) + \nonumber \\
&& -2 \hbar^2 g_k(x) g_k''(x) X'^2_k (x)  = 8m \left( En-U_k (x)\right) g_k^4(x)  X'^2_k(x)
\end{eqnarray}			
Eq. (14), when $\hbar \to 0$  becomes the $CHJE$, so that, $X_k(x)$ in this limit becomes the classical action $W_{C,k} (x)$.
Beyond the arc's ends, along the continuation of the arc itself, the quantum action has only the imaginary part $Y_k(x)$, which satisfies the same Eq. (11), but with $X_k(x) = 0$.

When a solution of Eq. (14) is known, by putting it and $Y_k(x)$ as computed from Eq. (13) into Eq. (10), and then by using Eq. (9), a complex solution of the SE along the caustics arc is obtained. By combining it and its conjugate, this solution can be put into the trigonometric WKB-like form [2-4]:
\begin{equation}
\psi _k(x)={A_k\over \sqrt{X_k' (x)\over g_k(x)}} {\rm Sin}\left({X_k(x)\over \hbar}+{\pi\over 4}\right) \quad ,
\end{equation}		
where $A_k$ is a constant. The Eq. (15) is the exact generalization of the well-known WKB representation of the semi classical wave function. Differently from this latter, it is valid also at the turning points where $U_k(x) = E$, i.e. at the arc's ends. Externally, on the arc's continuation at the left side with respect to the (oriented) arc itself, the solution of the SE has the form:
\begin{equation}
\psi_{k,l}(x)=B_{k,l}{\rm Exp}\left(Y_{k,l}(x)\right)\quad ,
\end{equation} 	
where $B_{k,l}$ is a constant and the suffix $l$ means left. An analogous expression holds for the solution on the right continuation of the arc. Obviously, the functions $Y_k(x)$ in the Eq. (16) have to be chosen to give the suitable behavior of the wave functions for large values of $|x|$. As detailed in [3], when for each arc the function (15), at the arc' ends, smoothly matches together with its first derivative with the solutions (16) found on the arc's left and right continuations, it is the restriction to the arc of the 2-dim wave function corresponding to the energy E, and the quantum action, the energy eigenvalues and the wave functions on the caustic's arcs are found. In this way, the problem to find the energy eigenvalues for the 2-dim system reduces to 1-dim problems on each caustic's arc. If one is instead interested to find the quantum action not simply on the caustic, but in the full region $\Omega_F$ too, and from it, the full 2-dim wave function there, one has to solve the Dirichlet's problem for the QHJE in $\Omega_F$, using as boundary values on $C_F$ the quantum actions previously calculated on the caustic's arcs.

In order to implement the program above, the procedure starts by choosing a tentative value for the energy E. Then a classical trajectory with this energy is computed, by integrating the Hamilton's equations for enough time to generate a well defined caustic, and choosing as initial conditions a point $x_V,y_V$ on the equipotential line $U(x, y) = E$. This point has therefore $\vec p = 0$ and in the generic case is a vertex of the caustic. The caustic itself is obtained by recording the zeroes of the external product of two independent solutions of the Jacobi equation for the trajectory [9, 10]. As known, the Jacobi equation gives the time evolution of the vector connecting two neighboring extremals (according to the Maupertuis principle, iso-energetic trajectories are the extremals of the classical reduced action [11]). As the caustic is the locus of the conjugated points to the initial point, these zeroes define the caustic [12]. Each caustic arc $k$ is then fitted by means of a smooth function $y=f_k(x)$ or $x=f_k(y)$, which is thereafter inserted into the Eq. (14). After that, the unidimensional quantization method presented in [3] is applied, by using for each arc as energy the same value E, and as 1-dim potential energy, the appropriate $U_k(x)$, as given by Eq. (8).  If it is not possible to construct in this way four regular wave functions along the caustic arcs and their continuations, the procedure is repeated with a different initial vertex. When no choice of this latter brings to regular wave functions, one has to try with a different energy, and so on.

Some results obtained in this way are reported in the table below, where the energy values computed by means of this method are compared with the corresponding ones obtained by diagonalization of the truncated Hamiltonian matrix. As seen from the table, the difference between the corresponding values is usually of few thousandths. For each state, the corresponding quantum numbers, i. e. the numbers of the nodal lines, are presented.
\begin{table}[!h]
\centering
\renewcommand{\arraystretch}{0.75}
\label{my-label}
\begin{tabular*}{0.70\textwidth}{|r|r|r|}
\hline
\parbox[t]{3.70cm}{Energy by trunc. \\Hamiltonian matrix} & \parbox[t]{3.70cm}{Energy by \\the present method} & \parbox[t]{3.70cm}{Quantum numbers\\ (m, n)} \\ \hline
1.04795 & 1.04795 & (0, 0)  \\ \hline
2.04496 & 2.04134 & (0, 1) \\ \hline
2.13613 & 2.13660 & (1, 0) \\ \hline
3.04143 & 3.04326 & (0, 2) \\ \hline
3.12687 & 3.12603 & (1, 1) \\ \hline
3.21568 & 3.21722 & (2, 0) \\ \hline
4.03723 & 4.04602 & (0, 3) \\ \hline
4.11580 & 4.11745 & (1, 2) \\ \hline
4.20074 & 4.20199 & (2, 1) \\ \hline
4.28659 & 4.28898 & (3, 0) \\ \hline
5.03219 & 5.03517 & (0, 4) \\ \hline
5.10257 & 5.10884 & (1, 3) \\ \hline
5.18266 & 5.18871 & (2, 2) \\ \hline
5.26680 & 5.26396 & (3, 1) \\ \hline
5.34893 & 5.34668 & (4, 0) \\ \hline
6.02600 & 6.04134 & (0, 5) \\ \hline
6.08666 & 6.09051 & (1, 4) \\ \hline
6.16106 & 6.16192 & (2, 3) \\ \hline
6.24230 & 6.24120 & (3, 2) \\ \hline
6.32531 & 6.32422 & (4, 1) \\ \hline
6.40268 & 6.40674 & (5, 0) \\ \hline
7.01807 & 7.04359 & (0, 6) \\ \hline
7.06720 & 7.06945 & (1, 5) \\ \hline
7.13551 & 7.13708 & (2, 4) \\ \hline
7.21297 & 7.21429 & (3, 3) \\ \hline
7.29506 & 7.29003 & (4, 2) \\ \hline
7.37667 & 7.37235 & (5, 1) \\ \hline
7.44811 & 7.44841 & (6, 0) \\ \hline
8.00716 & 8.03765 & (0, 7) \\ \hline
8.04281 & 8.03829 & (1, 6) \\ \hline
8.10566 & 8.10707 & (2, 5) \\ \hline
8.17860 & 8.17905 & (3, 4) \\ \hline
8.25864 & 8.24276 & (4, 3) \\ \hline
8.34128 & 8.33645 & (5, 2) \\ \hline
8.42139 & 8.41825 & (6, 1) \\ \hline
8.48549 & 8.48295 & (7, 0) \\ \hline
\end{tabular*}
\caption{The lowest energy eigenvalues for the Barbanis system, as computed by diagonalization of the truncated Hamiltonian matrix (left column), compared with those computed by the method presented in the paper. The quantum numbers ($m$, $n$) are also indicated. Parameters values as in Fig. [1], and $\hbar=  1$.}
\end{table}
For the wave functions, too, the method gives results in very good agreement with other approaches. For instance, the Fig. [4] is the graph of the potential $U_3(y)$, i.e. the effective unidimensional potential along the caustic's arc $k=3$ (the right arc in the Fig. [1]). In Fig. [5] is plotted the wave function calculated with this method on the same arc (continuous line), together with the corresponding wave function computed by means of diagonalization of the truncated Hamiltonian matrix (dotted line). The figures refers to the state (2, 2) with the energy $E= 5.18266$. Only the oscillating parts of the wave functions, between the two vertices, are plotted. As seen from the figure, the two curves are practically overlapped.

\begin{figure}[!h]
\includegraphics[scale=1]{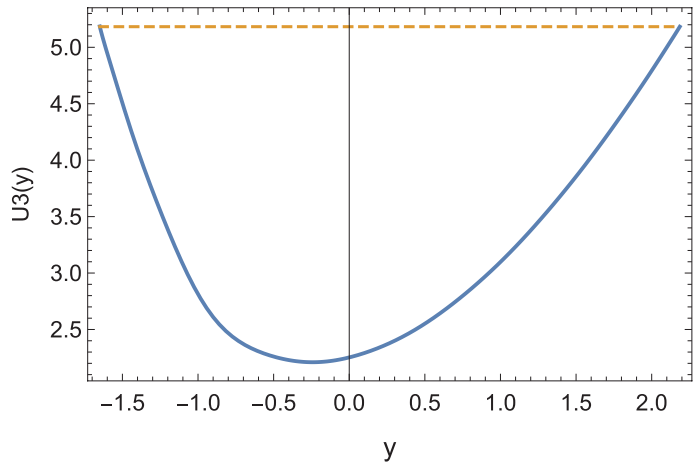}%
\caption{The graph of the potential $U_3(y)$, i.e. the effective unidimensional potential along the caustic's arc $k=3$ (the right arc in the Fig. [1]). The dashed line represents the total energy $E=5.18266$.}
\end{figure} 
\begin{figure}[!h]
\includegraphics[scale=1]{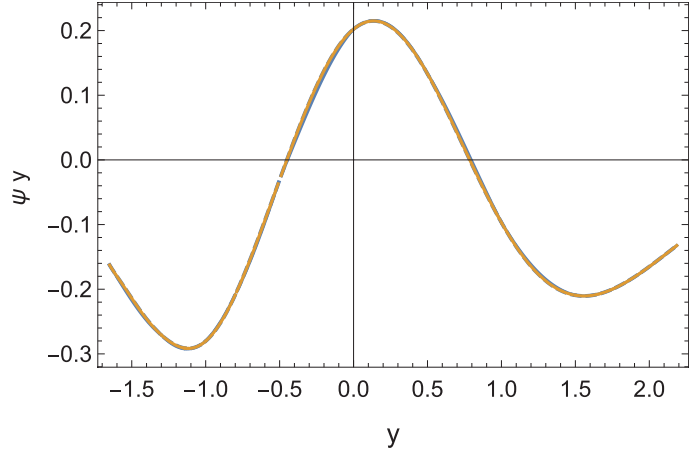}%
\caption{The wave functions calculated with the present method on the arc with $k = 3$, (continuous, yellow line), and the corresponding wave function computed by means of diagonalization of the truncated Hamiltonian matrix (dotted blue line). The figures refers to the state (2, 2) with the energy $E= 5.18266$. Only the oscillating parts of the wave functions, between the two vertices, are plotted. As seen from the figure, the two curves are practically overlapped. }
\end{figure} 

The condition discussed in [2 - 4] for the 1-dim. case, concerning the variation $\Delta X_k= X_k(x2)-X_k(x1)$ of the real part of the quantum action between the two turning points $x1$ and $x2$, applies to both the caustic's arcs coming out from each vertex. By indicating by $j$ and $k$ two such arcs, and with $\Delta X_j$ and  $\Delta X_k$ the respective variations between the arcs' ends, therefore one has:
\begin{eqnarray}
\Delta X_j &=& \left ( m+{1\over 2}\right) \nonumber \\
\Delta X_k &=& \left ( n+ {1\over 2}\right)
\end{eqnarray}    
where $m$ and $n$ are two integer that correspond to the number of nodes of the wave functions on the two arcs, respectively. These variations can also be written as line integrals of the derivatives of the quantum actions with respect to the argument, which are the functions that in the limit $\hbar \to 0$ generate the classical momenta [2 - 4].  In this limit, in terms of line integrals along the arcs, the equations (17) become:
\begin{eqnarray}
\int_j \vec p d \vec q &=& \hbar \pi \left(n+{1\over 2}\right)\nonumber\\
\int_k \vec p d\vec q &=& \hbar \pi \left(m+{1\over 2}\right)\nonumber\\
\end{eqnarray}
which, as demonstrated by Marcus [ 13 ]  are the projection of  the usual EBK semi classical quantization conditions [14], from the phase space to the configuration space.

In conclusion, from the previous analysis it follows that a value E of the energy is allowed in the quantum case for the systems under consideration, only if there exists a family of classical trajectories at that energy, such that the corresponding caustic can host in its arcs the complete oscillating part of four 1-dim wave functions with the same energy. By complete we mean the part of the wave function between the first and the last zeroes of its second derivative.

Finally, the results presented demonstrate that to every quantum wave function for a 2-dim non-separable but integrable Hamiltonian system, it corresponds a particular family of classical trajectories. Their link is given by the quantum action. Moreover, for these systems it is possible to compute, on the classical caustic's arcs, the quantum actions, the energy eigenvalues and the values of the wave functions, by integrating the one-dimensional QHJ equations on the arcs itself. If one is also interested to compute the quantum action inside and outside the region $\Omega_F$, and thereafter the wave functions, it is necessary to integrate the QHJE with the previously calculated values on the boundary. This will be the object of a future publication. If instead one is only interested to find the energy eigenvalues and the wave functions on the caustic's arcs, it is not necessary to solve the 1-dim QHJ equations, and one can simply solve the 1-dim Schr\"odinger equations on the arcs. For both the methods, the results are obviously the same. Extension of the method presented here to chaotic 2-dim systems and to more dimensions is presently under investigation.

\end{document}